\documentclass[12pt]{article}

\usepackage{amssymb}
\usepackage{epsfig,color}
\usepackage{pstricks,graphicx,epsfig,color,amssymb,amsmath,amscd}
\usepackage{cite}
\usepackage[]{graphicx}
\usepackage{makeidx}
\newcommand{\bi}{\bibitem}
\newcommand{\be}{\begin{eqnarray}}
\newcommand{\ee}{\end{eqnarray}}
\newcommand{\rar}{\rightarrow}

\usepackage[]{caption}
\begin{document}

\begin{titlepage}
\title{ Explosive phenomena in modified gravity} 
\author{E.V. Arbuzova$^a$, A.D. Dolgov$^{b,c,d}$}

\maketitle
\begin{center}
$^{a}$Department of Higher Mathematics, University "Dubna", 141980 Dubna, Russia\\
$^{b}$Dipartimento di Fisica, Universit\`a degli Studi di Ferrara, 
I-44100 Ferrara, Italy \\
$^{c}$Istituto Nazionale di Fisica Nucleare, Sezione di Ferrara, 
I-44100 Ferrara, Italy \\
$^{d}$Institute of Theoretical and Experimental Physics, 113259 Moscow, Russia 
\end{center}

\begin{abstract}
Observational manifestations of some models of modified gravity, which
have been suggested to explain the accelerated cosmological
expansion,  are analyzed for gravitating systems with time dependent 
mass density. It is shown that if the mass density rises with time, the 
system evolves to the singular state with infinite curvature scalar. The 
corresponding characteristic time is typically much shorter than the 
cosmological time.
\end{abstract}

\end{titlepage}

Contemporary astronomical data strongly indicate 
that at the present epoch the universe expands 
with acceleration.
A possible way to explain this accelerated expansion is to assume that
there is a new component in the cosmological energy density, the so
called dark energy. The latter can be either a small vacuum energy,
which is identical to the cosmological constant, or the energy density
associated with an unknown, presumably scalar field, which slowly
varies in the course of the cosmological evolution.

A competing possibility to create 
cosmological acceleration is to modify gravity itself introducing
additional terms into the usual action of General
Relativity~\cite{grav-mdf}; for recent reviews see~\cite{NojOd,App-Bat-Star}. 
To this purpose the models with the following action
were considered:
\be
S = \frac{m_{Pl}^2}{16\pi} \int d^4 x \sqrt{-g} f(R)+S_m, 
\label{A1}
\ee 
where $m_{Pl}= 1. 2 2\cdot 10^{19}$ GeV is the Planck mass, 
$R$ is the scalar curvature, and $S_m$ is 
the matter action. In the usual Einstein gravity function $f(R)$ has the 
form $f(R)=R$, in the modified gravity $f(R)$ acquires an additional term:  
\be
f(R)=R+F(R),
\label{A2}
\ee 
which changes gravity at large distances and is responsible for
cosmological acceleration. In the pioneering papers~\cite{grav-mdf}
function $F(R)$ at small $R$ behaves as:
\be 
F(R) = - \frac{\mu^4}{R}\,,
\label{F}
\ee 
where $\mu$ is a small parameter with dimension of mass.
However, as it was shown in ref.~\cite{DolgKaw}, such a choice of $F(R)$ 
leads to a strong exponential instability near massive objects
and so the usual gravitational fields would be drastically distorted. 
An attempt to cure this ill-behavior by adding to the action
$gR^2$-term~\cite{noj-od-R2} was only partially successful. It 
could terminate the instability with reasonably small coefficient $g$
for sufficiently dense objects with $\rho > 1\,\, {\rm g/cm}^3$,
while for the objects with smaller mass density the coefficient $g$
would be too large and incompatible with the existing bound on the
$R^2$-gravity. We will discuss this problem in a more detailed paper
which is under preparation.

A choice of $F(R)$, which leads to an accelerated cosmological
expansion and is devoid of the above mentioned instability and of some other
problems was suggested in several papers~\cite{Starob, HuSaw, ApplBatt,NojOdin}.
In the present work we examine a very interesting model of modified 
gravity with $F(R)$ function suggested  in ref.~\cite{Starob} : 
\be
F(R)= \lambda R_0 \,\left[ {\left(1+ \frac{R^2}{R_0^2}\right)^{-n}} - 1 \right]\,.
\label{F-AAS}
\ee
Here constant $\lambda $ is chosen to be positive to produce an
accelerated cosmological expansion, $n $ is a positive integer, and 
$R_0$ is a constant with dimension of the curvature scalar.  
The latter is assumed to be of the order of
the present day average curvature of the universe, i.e. 
$R_0 \sim 1/t^2_U$, where $t_U \approx 4\cdot 10^{17}$ sec is the universe age.

The corresponding equations of motion have the form
\be
\left( 1 + F'\right) R_{\mu\nu} -\frac{1}{2}\left( R + F\right)g_{\mu\nu}
+ \left( g_{\mu\nu} D_\alpha D^\alpha - D_\mu D_\nu \right) F'  = 
\frac{8\pi T^{(m)}_{\mu\nu}}{m_{Pl}^2}\,,
\label{eq-of-mot}
\ee 
where $F' = dF/dR$, $D_\mu$ is the covariant derivative,  and 
$T^{(m)}_{\mu\nu}$ is the energy-momentum tensor of matter.

By taking trace over $\mu $ and $\nu $ in eq. (\ref{eq-of-mot}) we obtain the equation
of motion which contains only the curvature scalar $R$  and the trace
of the energy-momentum tensor of matter: 
\be
3 D^2 F' -R + R F' - 2F = T\,,
\label{D2-R}
\ee
where $T = 8\pi T_\mu^\mu /m_{Pl}^2$. Note that our sign convention is
different from that of paper~\cite{Starob} and is the same as in 
ref.~\cite{DolgKaw}.

Cosmology with gravitational action (\ref{F-AAS}), as well as some other
cosmological scenarios with modified gravity were 
critically analyzed in recent paper~\cite{App-Bat-Star}. It was 
shown that, taken literally, the models
suffer from several serious problems. Though the instability of 
ref.~\cite{DolgKaw} was eliminated, still there remain some other
types of singular behavior.  In particular, there exists the past
singularity, when $R\rar \infty$ at some finite time in the past.
It was argued that the problem can be solved by an addition 
to the action of $R^2$-term with sufficiently small coefficient
allowed by the present observational data. 

The singularity similar to that considered in the present work was first 
noticed in ref.~\cite{appl-bat-08} in the case of cosmological evolution back to the past.
In a sense the future singularity considered here is the time reversal
of the past singularity on the quoted paper. So mathematically both singularities are 
quite similar, despite of some difference due to effects of the universe contraction
(when one goes backward in time). The Hubble anti-friction favors the approach to 
the singularity in the contracting universe. 
However, despite mathematical similarities there is an important difference between 
the two systems. According the
ref.~\cite{appl-bat-08}, the singularity may be avoided with a certain range of initial
conditions. In our case singularity emerges for any initial conditions.

In ref.~\cite{sami,frolov} it was
argued that infinite $R$ singularity could arise in the future, unless
the initial conditions for $R$ are not  fine-tuned.  This is similar to the cosmological
situation of ref.~\cite{appl-bat-08}. The systems
considered in these works are different from that discussed here and
the singularity of the quoted papers appears only for 
certain initial conditions, while in our case, as we have already mentioned above,
 the singularity arises for an arbitrary initial state. 
All the singularities can  be eliminated by an addition of $R^2$-term to the
action and we study the effects of this term below.

In what follows we will consider a different physical situation than
those discussed in the above mentioned references. Namely we study
behavior of astronomical objects with mass density which rises with
time and show that curvature, $R$, reaches infinitely large value
during the time interval which is very short in comparison with the
cosmological time scale. This singularity cannot be eliminated by
fine-tuning of the initial conditions. An addition of $R^2$-term could
prevent from the singular behavior but at expense of quite large
values of $n$ which may be at odds with the standard cosmological
evolution. 

We study objects with mass density which is much
larger than the cosmological one, $\rho_m \gg \rho_c $.  The
cosmological energy density at the present time is 
$\rho_c \approx 10^{-29}\,{\rm g/cm}^3$, while matter density of, say,
a dust cloud in a galaxy could be about $\rho_m \sim 10^{-24}{\rm g/cm}^3$.   
Since the magnitude of the curvature scalar is proportional to the
mass density of a nonrelativistic system, we find $R \gg R_0 $. In
this limit: 
\be
F(R) \approx -\lambda R_0 \left[ 1 -\left(\frac{R_0}{R}\right)^{2n} \right] \,.
\label{F-large-R}
\ee

Let us start from the initial state in which 
modified gravity around or inside some massive objects is not
much different from the usual Einstein (Newtonian) gravity and
correspondingly $R \approx - T$, as can be seen from the 
normal Einstein equations.

We analyze temporary evolution of  
solutions of eq. (\ref{D2-R}) for the gravitational field of some 
massive object with time varying density. 
We assume that the gravitational
field of this object is weak, as is usually the case. Correspondingly
the background metric is approximately flat and  
the covariant derivatives can be replaced by the usual flat space ones. 
Hence:
\be
D^2 F' = (\partial^2_t - \Delta) F' = F'' (\partial^2_t - \Delta) R + 
F''' [(\dot R)^2 - (\nabla R)^2 ]
\label{D2-F'}\,,
\ee
where $\Delta$ is the usual Laplacian, and $\nabla$ is the gradient.

Substituting expression (\ref{F-large-R})
for $F(R)$ at large $R$ into eq.~(\ref{D2-R}),
we obtain:
\be
\nonumber
(\partial^2_t - \Delta) R -(2n+2) \frac{\dot R^2 - (\nabla R)^2}{R} +
\frac{R^2}{3n(2n+1)} \left(\frac{R^{2n}}{R_0^{2n}} -(n+1) \right) \\
\qquad -\frac{R^{2n+2}}{6n(2n+1)\lambda R_0^{2n+1}} (R + T)=0\,.  
\label{eq-for-R}
\ee
However, due to the presence of the nonlinear terms containing
derivatives, this equation is difficult to analyze and we instead
study the equation for $F'(R)$ and express $R$
through $F'$ using:
\be
F' = - 2n\lambda \left(\frac{R_0}{R}\right)^{2n+1} \,.
\label{F'-of-R}
\ee
Notice that infinite $R$ corresponds to $F' =0$ and if 
$F'$ reaches zero, it would mean that $R$ becomes infinitly large.

Let us introduce the new notation $w = - F'$. Equation (\ref{eq-for-R}) for $w$ 
takes the simple form describing an unharmonic oscillator:
\be
(\partial^2_t - \Delta) w  + U'(w) = 0\,.
\label{eq-for-w}
\ee
Potential $U(w)$ is equal to:
\be
U(w) = \frac{1}{3}\left( T - 2\lambda R_0\right) w + 
\frac{R_0}{3} \left[ \frac{q^\nu}{2n\nu} w^{2n\nu}+ \left(q^\nu
    +\frac{2\lambda}{q^{2n\nu} } \right) \,\frac{w^{1+2n\nu}}{1+2n\nu}\right]\,,
\label{U-of-w}
\ee
where $\nu = 1/(2n+1)$, $q= 2n\lambda$, and in
eq. (\ref{eq-for-w}) $U'(w)=dU/dw$. 
It is useful to remember that $T\gg R_0$. Their ratio is about 
$T/R_0 \sim \rho_m/\rho_c \gg 1$ and hence $w\ll 1$. 
Thus the first term in square brackets in  eq. (\ref{U-of-w}) dominates. 
Potential $U$ would depend upon time, if the mass
density of the object under scrutiny changes with time, $T=T(t)$.

If only the dominant terms are retained in equations 
(\ref{eq-for-w}), (\ref{U-of-w})
and if the space derivatives are neglected, equation
(\ref{eq-for-w}) simplifies to:
\be
\ddot w + T/3 - \frac{q^\nu (-R_0)}{3w^\nu}=0\,.
\label{eq-w-simple}
\ee 

It is convenient to introduce dimensionless quantities:
\be
t = \gamma \tau,\,\,\, w = \beta z\,,
\label{dimless}
\ee
where $\beta$ and $\gamma$ are so chosen that the equation for $z$
becomes very simple:
\be
z'' - z^{-\nu} + (1+\kappa \tau) = 0\,.
\label{eq-for-z}
\ee
Here prime means differentiation with respect to $\tau$ and the trace
of the energy-momentum tensor of matter is parametrized as:
\be
T(t) = T_0 (1 + \kappa \tau)\,. 
\label{T-of-t}
\ee
Constants $\gamma$ and $\beta$ are equal to
\be
\gamma^2 = \frac{3q}{(-R_0)} \left(-\frac{R_0}{T_0}\right)^{2(n+1)}\,,\\
\beta = \gamma^2T_0/3 = q \left(-\frac{R_0}{T_0}\right)^{2n+1}\,.
\label{gamma-beta}
\ee
Thus $\beta$ is a small dimensionless number, and $\gamma$ has dimension
of time. It is essential that $\gamma$, which determines
characteristic time scale, may be much shorter than the universe age,
$t_U$, due to the small factor $(R_0/T_0)^{n+1}$. Assuming that
$3q \sim 1$ and $R_0 \sim 1/t_U^2$, we find for $n=2$ and 
$\rho_m = 10^{-24}\,{\rm g/cm}^3$: $\gamma \approx 400$ sec.
It would be much smaller for larger $n$ or $\rho_m$. For example if 
$n=3$ and the same $\rho_m$ we find $\gamma = 0.004$ sec.

In the case of constant $T$ ($\kappa =0$) or very slowly varying $T$
($\kappa \ll 1$) the solution of eq. (\ref{eq-for-z}) 
is evident. If the initial values $z(0)$ and $z'(0)$ are sufficiently small, 
$z(\tau )$ oscillates near the minimum of the potential, which is
situated at
\be 
z_{min} = (1+\kappa \tau )^{-1/\nu} \,.
\label{z-min}
\ee
If by some reason 
the magnitude of $z(0)$ takes a sufficiently large value, 
$z> (1-\nu)^{1/\nu}$,  such that  potential 
\be
U(z)=z-z^{1-\nu }/(1-\nu ) \,.
\label{U-of-z}
\ee
becomes positive, evidently at some stage $z(\tau )$ would 
overjump potential $U(z)$ which is equal to zero at $z=0$ 
("the waves are cresting over"). In other words, 
$z(\tau )$ would reach zero, which corresponds to infinite $R$, and so
the singularity can be reached in finite time. Analogous situation can be
realized if the initial velocity, $z'(0)$, is  sufficiently large. 

The singularity can be also reached in finite time even if 
$z$ was initially situated at the minimum of the potential
and the initial velocity was zero. It would take place
if $\kappa$ is positive, i.e. the energy density rises with time.
The motion of $z_{min}$ to zero and simultaneous diminishing 
of the depth of the potential well make it easier for $z(\tau)$ to reach
zero. On the other hand,  it is
not evident that for decreasing energy density 
$z(\tau)$ initially resting at the minimum of $U(z)$ could reach
zero, most probably it would not.

We have solved equation (\ref{eq-for-z}) numerically and
found that indeed the singularity, $R\rar \infty$, is reached in finite time
for rising $T(\tau)$ under quite general conditions. The solution for
$n=2$, $\kappa = 0.01$, and $\rho_m/\rho_c = 10^5$ is presented in
Fig.~1, where the ratio $z(\tau)/z_{min}(\tau)$ (left) and 
functions $z(\tau)$ and $z_{min} (\tau)$ separately (right) 
are depicted.
The initial conditions are taken as $z(0) =1$ and $z'(0) =0$.

In Figs.~2 and 3 the same quantities are presented for $n=3 $ and
$n=4$ respectively. It is clearly seen that $z(\tau)$ reaches
zero after a finite number of oscillations around $z_{min}(\tau)$. 
When $z_{min} (\tau)$ shifts to smaller values, function
$z(\tau)$ initially remains behind but when the displacement from the
equilibrium point becomes large enough, $z(\tau)$ started to run after it
with an increasing speed, then overtakes the position  of the
minimum, and oscillates back. After a few oscillations the retarded
position of $z(\tau)$ happens to be above the point $z_0(\tau)$, where the
potential is zero. It is essential that the position of this
point moves to smaller values with
rising $T(\tau)$. Because of that it is easier to overjump the
potential at $z=0$. It is intriguing that the magnitude of the ratio
$z/z_{min}$ is approximately equal to 3 in the last maximum before the
singular point $z=0$ is reached. We have not found an explanation for that.

\begin{figure}
\begin{center}
     \includegraphics[scale=0.8]{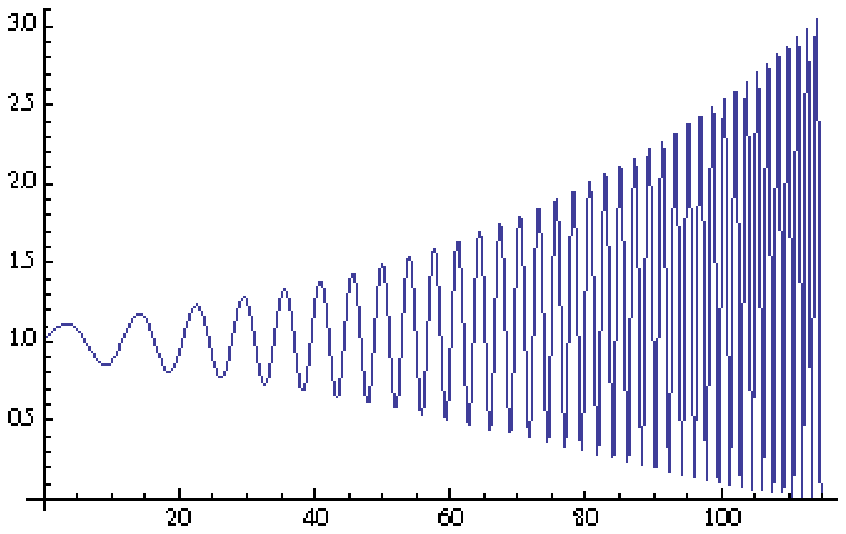} \hspace{.7cm}
    \includegraphics[scale=0.8]{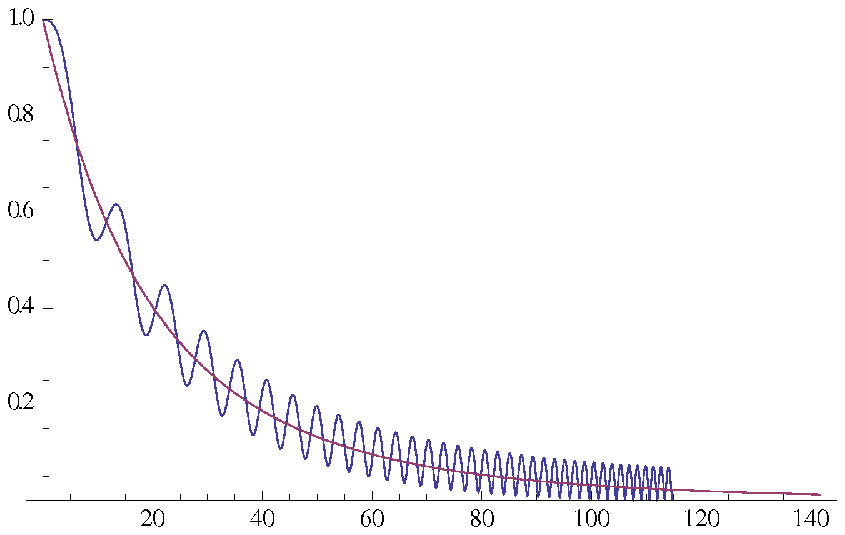}
        \caption{
Ratio $z(\tau)/z_{min}(\tau)$ (left) 
and  functions $z(\tau)$ and $z_{min}(\tau)$ (right) for
$n=2$, $\kappa = 0.01$, $\rho_m/\rho_c = 10^5$. 
    \label{f-12}}
\end{center}
\end{figure}

\begin{figure}
\begin{center}
    \includegraphics[scale=0.8]{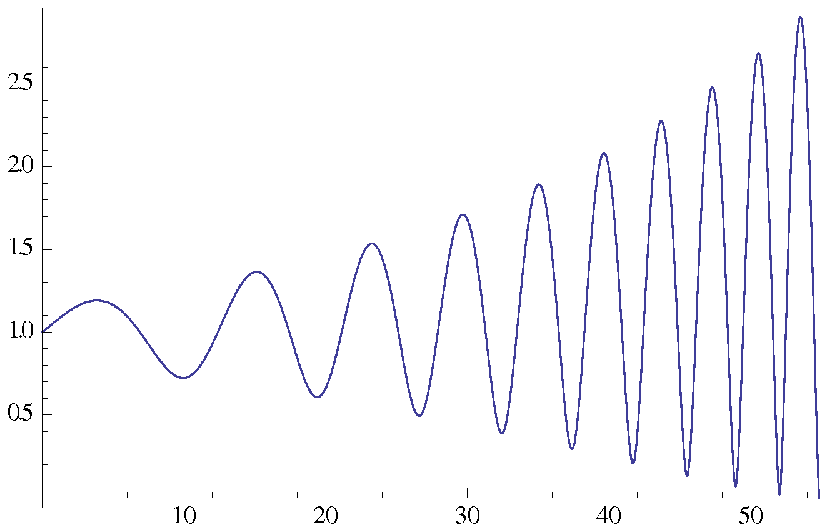} \hspace{0.7cm} 
    \includegraphics[scale=0.8]{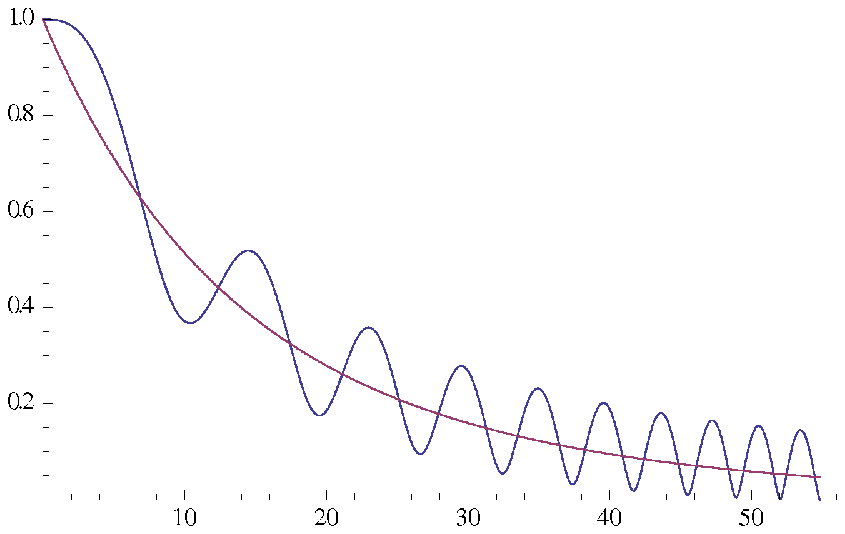}
 \caption{
Ratio $z(\tau)/z_{min}(\tau)$ (left) 
and  functions $z(\tau)$ and $z_{min}(\tau)$ (right) for
$n=3$, $\kappa = 0.01$, $\rho_m/\rho_c = 10^5$.}
    \label{f-34}
\end{center}
\end{figure}

\begin{figure}
\begin{center}
    \includegraphics[scale=0.8]{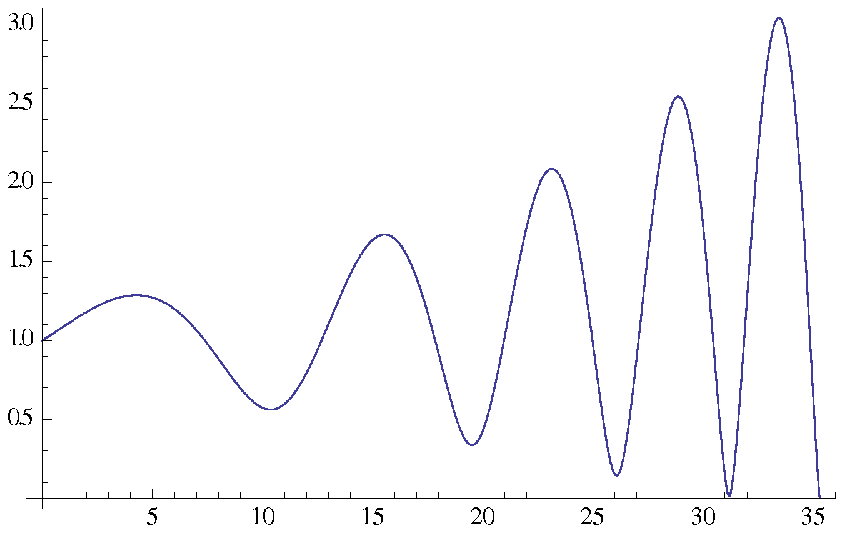} \hspace{0.7cm} 
    \includegraphics[scale=0.8]{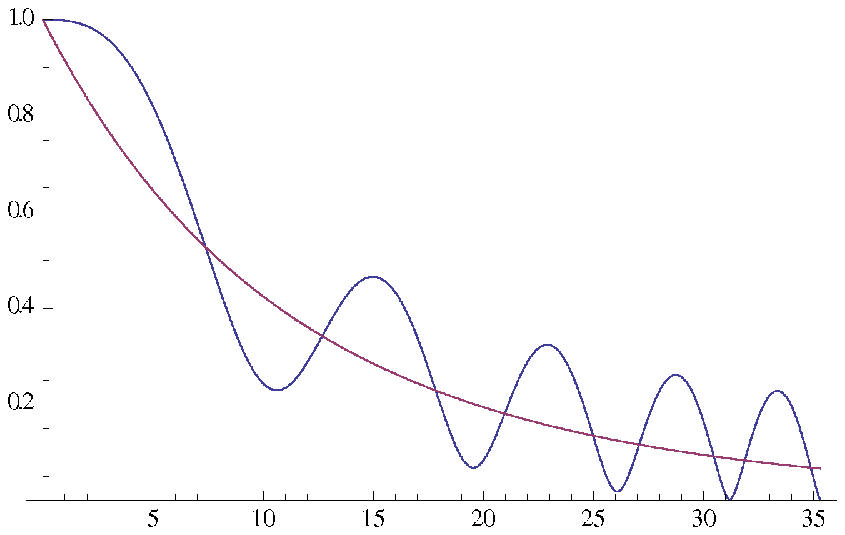}
\caption{Ratio $z(\tau)/z_{min}(\tau)$ (left) 
and  functions $z(\tau)$ and $z_{min}(\tau)$ (right) for
$n=4$, $\kappa = 0.01$, $\rho_m/\rho_c = 10^5$.}
    \label{f-56}
\end{center}
\end{figure}

In terms of physical time, $t$, the evolution of the energy density can
be presented as $T(t) = T_0 (1+ t/t_{ch})$, where  $t_{ch}$  is the characteristic 
time of the variation. 
Coefficient $\kappa$ in equation (\ref{T-of-t}) is expressed
through $t_{ch}$ as:
\be
\kappa = \gamma/t_{ch}\,.
\label{kappa}
\ee 
Thus the presented in Fig.~1 case of $\kappa = 0.01$, $n=2$, and
$\rho_m/\rho_c =10^5$ corresponds to $t_{ch} =  4\cdot 10^4$ sec,
while $n=3$ (Fig.~2) corresponds to $t_{ch} = 0.4$ sec. The characteristic time
of the density variation can be estimated as $t_{ch} \sim d/v$, where
$d$ is the size of the system and $v$ is the velocity of the
constituent particles in the process of the collapse of the cloud or
in the collision of the clouds. 
In the first case the velocity would be quite low and the characteristic time is expected
to be close to the Newton free-fall time but in the case of colliding clouds the velocities
are typically galactic ones, about 300 km/sec. The velocity may be even larger
at the collision of the supenova ejecta with galactic or intergalactic clouds.

It is more informative to act another way around, namely to estimate
$\kappa$ knowing size, $d$, of the object with changing mass density   
or sizes of the colliding objects:
\be
\kappa = \frac{\gamma v}{d} \,.
\label{kappa-of-d}
\ee
For $n>2$ and astronomically large clouds one should expect
$\kappa \ll 1$. For very small $\kappa$ our numerical
calculations with quickly oscillating functions are not reliable but
it seems natural to expect that the system would reach singularity
according to the analysis presented above.

As it is seen from the numerical calculations, the singularity is reached
when $t\sim t_{ch}$. This is much shorter than the cosmological time
for clouds of denser matter in galaxies or a collapsing cloud forming
a star or another denser body.

In the analysis of eq.~(\ref{eq-for-w}) the spatial derivates have been
neglected. At first sight, the account of these terms could inhibit formation
of singularity, as e.g. happens in the process of structure formation
due to gravitational (Jeans) instability. However the situation is
opposite here and the inhomogeneities stimulate singularity
formation. Indeed, the effect of inhomogeneities can be described by
an appearance of the term $w/d^2$ in eq. (\ref{eq-for-w}) with
positive coefficient. Such a term is equivalent to an addition of
an extra attractive force pushing $w$ or $z$ to zero, i.e. to 
$R\rar\infty$. 

There are several possible cases when the conditions leading to singularity
can be realized: collapse of gas cloud leading finally to star
formation, collision of two gas clouds in a galaxy, stellar ejecta
colliding with interstellar or intergalactic matter, and many others. 
From the calculational point of view such processes can be either
adiabatic, when the mass density changes slowly (this is the case
analyzed above) or fast, when the mass density changes instantly in an
explosive way. Seemingly the latter would result in a faster approach
to singularity. This case  will be analyzed elsewhere.

If $R$ becomes large, the approximation of flat space-time would be
invalid and the derivatives in the equations of motion should be
changed into covariant ones. We have not analyzed if in this case
the approach to singularity is terminated. However, even if it is
terminated, it takes place at high curvatures when gravity becomes
strongly different from the Newtonian one.

Another possible way to avoid singularity is to introduce $R^2$-terms into the
gravitational action:
\be
\delta F(R) = -R^2/ 6m^2\,,
\label{delta-F}
\ee
where $m$ is a constant parameter with dimension of mass. 

With such an extra term in the gravitational action it becomes impossible
to express analytically $R$ through $F'(R)$. So one needs to work with
the equation of motion for $R$. In the homogeneous case and in the limit
of large ratio $R/R_0$ equation (\ref{eq-for-R}) is modified as
\be
\left[ 1-\frac{R^{2n+2}}{6\lambda n(2n+1) R_0^{2n+1} m^2 }\right]\,\ddot R 
- (2n+2) \,\frac{\dot R^2}{R} -
\frac{R^{2n+2} (R+T)}{6\lambda n (2n+1) R_0^{2n+1}} = 0 \,.
\label{eq-for-R-mdf}
\ee

To analyze eq. (\ref{eq-for-R-mdf}) let us introduce, as is done
above, dimensionless curvature and time:
\be
y = -\frac{R}{T_0}\,,\,\,\,\,
\tau_1 =  t \left[-\frac{T_0^{2n+2}} { 6\lambda n (2n+1) R_0^{2n+1}}
\right]^{1/2} \,.
\label{y-x}
\ee
Correspondingly eq.~(\ref{eq-for-R-mdf}) is transformed into:
\be
\left(1 + g y^{2n+2} \right) y'' - 2(n+1)\,\frac{(y')^2}{y} + 
y^{2n+2} \left[y - (1 + \kappa_1 \tau_1)\right] = 0 \,,
\label{y-2prime}
\ee
where prime means derivative with respect to $\tau_1$ and
\be
g = -\frac{T_0^{2n+2}}{6\lambda n (2n+1) m^2 R_0^{2n+1}}>0 \,.
\label{g}
\ee
For very large $m$, or small $g$,
when the second term in the coefficient of the second derivatives in eqs. (\ref{eq-for-R-mdf}) 
and (\ref{y-2prime}) can be neglected,  
the numerical solution demonstrates that $R$ would reach
infinity  in finite time in accordance with the results presented above.
Nonzero $g$ would terminate the unbounded rise of $R$. To avoid too
large deviation of $R$ from the usual gravity coefficient $g$ should be larger than
or of the order of unity. Notice that the factor $(1 + g y^{2n+2})$ is always non-zero
because $g>0$.

Keeping in mind the bound on $m>10^{-2.5}$ eV, which follows from the
laboratory tests of gravity~\cite{grav-test}, we find $n\geq 6$,
demanding that the gravity of objects with $\rho\sim 10^{-24}\,{\rm g/cm}^3$
is not noticeably distorted. In ref.~\cite{App-Bat-Star} a stronger
bound is presented, $m\gg 10^5$ GeV. If this is the case, then
$n\geq 9$. A natural value is $m \sim m_{Pl}$ and correspondingly
$ n \geq 12$. For smaller values of $T_0$ the bounds on $n$ are
noticeably stronger.

As follows from eq. (\ref{y-2prime}), the frequency of small oscillations
of $y$ around $y_0 = 1+\kappa_1 \tau_1$ in dimensionless time  $\tau_1$ is
\be
\omega_\tau^2  = \frac{1}{g}\,\frac{g y_0^{2n+2}}{1+g y_0^{2n+2}} \leq \frac{1}{g}
\label{omega-tau}
\ee
It means that in physical time the frequency would be 
\be
\omega \sim \frac{1}{t_U} \left(\frac{T_0}{R_0}\right)^{n+1}
\frac{y_0^{n+1}}{\sqrt{1+g y_0^{2n+2}}} \leq m\,.
\label{omega}
\ee
In particular, for $n= 5$ and for a galactic gas cloud with $T_0/R_0 = 10^5$, 
the oscillation frequency would be 
$10^{12}\,\, {\rm Hz} \approx 10^{-3}$ eV.  Higher density objects
e.g. those with $\rho =1\,\, {\rm g/cm}^3$ would oscillate with much higher
frequency, saturating bound (\ref{omega}), i.e. $\omega \sim m$.
All kind of  particles with masses smaller than $m$
might be created by such oscillating field.

On the other hand, as we have seen above, for
denser objects the variation of $T$ in terms of $\tau$ or $\tau_1$ is
very slow because of very small $\kappa$. As a result the amplitude of the 
oscillations around the equilibrium point would be also small and
possibly such oscillations are of no danger from the observational point
of view. Still it is possible that there might be intermediate cases
when the oscillations would lead to observable phenomena.

Thus we have shown that the impact of the considered above versions 
of modified gravity on the systems with time dependent mass density 
in the contemporary universe could be catastrophic, leading to 
the singularity $R\rar \infty$ during finite time in the future.
This time is typically much shorter than the cosmological one. 
The problem can be fixed by the $R^2$-term if the power $n$
is sufficiently large, $n\geq 6$ (or maybe $n\geq 9$). So either the
versions of the theory with large $n$ or theories with another form 
of $F(R)$ should be considered. 
A more exciting possibility is that
the explosive phenomena predicted by modified gravity are observed in
the sky.


\end{document}